\documentclass[final,3p,times,onecolumn]{elsarticle}

\usepackage[english]{babel}
\usepackage{lineno,hyperref}
\modulolinenumbers[5]

\usepackage{multicol}
\usepackage{multirow}
\usepackage{url}
\usepackage{booktabs}
\usepackage{color}
\usepackage{textcomp}
\usepackage{amsmath}
\usepackage{amssymb}
\usepackage{xcolor}
\usepackage{algorithm}
\usepackage[noend]{algorithmic}
\usepackage{natbib}
\usepackage{subcaption}

\definecolor{mygreen}{HTML}{0f5f11}
\definecolor{amaranth}{rgb}{0.9, 0.17, 0.31}

\addto\extrasenglish{%

}

\journal{arXiv}

\begin{document}
\sloppy
\tolerance = 999

\begin{frontmatter}

\title{Implementing simple spectral denoising for environmental audio recordings}

\author[icmc]{F\'abio Felix Dias\corref{cor1}}
\ead{f\_diasfabio@usp.br}

\author[icmc]{Moacir Antonelli Ponti}

\author[icmc,ucc]{Rosane Minghim}

\cortext[cor1]{Corresponding author}
\address[icmc]{Instituto de Ci\^{e}ncias Matem\'{a}ticas e de Computa\c{c}{\~a}o, University of S{\~a}o Paulo \\ Av. Trabalhador S{\~a}o-carlense, 400, S{\~a}o Carlos, SP, Brazil, 13566-590}
\address[ucc]{School of Computer Science and Information Technology, University College Cork, Ireland}

\begin{abstract}
This technical report details changes applied to a noise filter to facilitate its application and improve its results.
The filter is applied to denoise natural sounds recorded in the wild and to generate an acoustic index used in soundscape analysis.
\end{abstract}


\end{frontmatter}


\section{Introduction}
\label{sec:introduction}

Signal distortions and noise are often present in real-life audio acquisition. In environmental audio recordings, background noise is even more evident due to the capture in the wild. 
When it comes to using the recordings as input to recognition systems, noise and distortions can impair accurate ecological interpretations of acoustic features. 

In ecological analysis via audio data, for example, as presented by~\citet{Dias:2021,Droge:2021,Gan:2020,Hilasaca:2021b,Scarpelli:2021}, acoustic features have an important role in the task of summarizing, representing, visualizing, and analyzing soundscapes.
Therefore, removing noise from an audio signal is an important pre-processing step to extract and analyze proper acoustic features related to soundscape dynamics.
Moreover, techniques as source separation are also used to help that analysis~\cite{Lin:2017,Lin:2020}.

In this report, we describe the denoising procedure proposed by~\citet{Towsey:2017}. We also aim at improving its performance and simplicity through the application of a simple algorithm to generate the \emph{noise profile} and the \emph{threshold} used in the algorithm. 

\section{Basic concepts}
\label{sec:theory}

In this section, we describe the original filter implementation, the threshold selection approach, and a statistical tool used to identify and remove outliers.

\subsection{Spectral filtering}

There are several filters available that are designed to be more general or serve specific denoising purposes. 
Among those techniques, the subtraction methods are considered to be straightforward. In particular, the subtraction using mean, median, or mode are often employed~\cite{Towsey:2017}.
In this report, we present a filter developed by~\citet{Towsey:2017}, which uses modal noise subtraction and assumes that the signal is generated by the addition of acoustic events energy with a Gaussian background noise.
We focused on a filter in the frequency domain, but a similar process can be used on the time domain.

\citet{Towsey:2017} created a variation of the \emph{adaptive level equalisation} method~\cite{Lamel:1981}, which assumes an additive Gaussian model, described in~\autoref{al:filter}.
First, the algorithm uses histograms to identify the mode value of each spectrogram band frequency and generate a noise profile of the spectrum.
In the next step, a smoothed profile is subtracted from the spectrogram rows and negative values are truncated to zero.
\citet{Towsey:2017} implemented an additional step that preserves complex events such as bird vocalizations.
They slide a window throughout the filtered spectrogram, calculating the mean and the minimum values of the region under the window.
If the mean is lower than a user-defined threshold, the minimum value is assigned to the spectrogram position where the window is centered. 

\begin{algorithm}[!t]
  \caption{Noise removal from spectrograms proposed by~\citet{Towsey:2017}}
  \label{al:filter}
  \begin{algorithmic}[1]
   \STATE \textbf{Input} signal $\mathbf{s}$ and threshold $\theta$
   \STATE $Sxx \leftarrow \text{stft}(\mathbf{s}, window = Hamming, size = 512, hop = 256)$   
   \STATE $S \leftarrow \text{abs}(Sxx)$ \\  
   \COMMENT{\textbf{Step A}}
   \STATE $profile \leftarrow \emptyset $   
   \FOR{ $i \leftarrow 1$ \textbf{to} $\#rows(S)$ } 
      \STATE $h \leftarrow \text{histogram}(S_i, bins = 100)$
      \STATE $h \leftarrow \text{moving\_average}(h, window = 5)$
      \STATE $M \leftarrow bin_j$ with max $h_j$
      \IF{ $j > 95$ }
        \STATE $M \leftarrow bin_{95}$
      \ENDIF      
      \STATE Insert $M$ in the $profile$
  \ENDFOR
  \COMMENT{\textbf{Step B}}
  \STATE $profile \leftarrow \text{moving\_average}(profile, window = 5)$
  \FOR{ $i \leftarrow 1$ \textbf{to} \#rows($S$) }
      \FOR{ $j \leftarrow 1$ \textbf{to} \#cols($S$) } 
        \STATE  $S_{ij} \leftarrow S_{ij} - profile_i$
        \IF{ $S_{ij} < 0$ }
          \STATE $S_{ij} \leftarrow 0$
        \ENDIF              
      \ENDFOR
  \ENDFOR
  \COMMENT{\textbf{Step C (additional)}}
  \STATE Centered in each $S_{ij}$ position, slide a $w_{9\times3}$ window throughout $S$  
  \STATE and calculate $avg_{ij}$ and $min_{ij}$ of the area under $w$.
  \IF{ $avg_{ij} < \theta$ }
    \STATE $S_{ij} \leftarrow min_{ij}$
  \ENDIF      
  \STATE  
  \STATE \textbf{Output} abs(istft($S\circ Sxx$)) \COMMENT{$\circ$ is the element-wise product}
  \end{algorithmic}  
\end{algorithm}

\subsection{Threshold selection}
\label{sec:otsu}

A recurring task in image processing and analysis is the definition of a threshold to separate foreground from the image background~\cite{Otsu:1979} that can be applied to image binarization and segmentation even as a post-processing step~\cite{ponti2012segmentation}.
\citet{Otsu:1979} developed a non-parametric and unsupervised approach to encounter a suitable threshold for these tasks.
Assuming the data is bimodal, i.e., it is distributed so that there are two main distinct parts or classes (in images usually foreground and background), the technique attempts to maximize the modal/class separability, evaluating the between-class variance ($\sigma_B^2$), iterating over the range values to encounter the better threshold $k$, by:
\begin{equation*}
    \max_{1 \le k < L} \sigma_B^2 (k), 
\end{equation*}
\noindent where $L = 255$ in a gray-scale image scenario.
Considering the values of the image histogram as $p_i$, it is possible to define the probability of $C_0$ (class before the threshold $k$) as $\omega(k) = \sum_{i = 1}^k p_i$, and the mean of that class as $\mu(k) = \sum_{i = 1}^k ip_i$.
The total mean can be calculated as $\mu_T = \sum_{i = 1}^L ip_i$.
In that way, the between-class variance could be expressed as:

\begin{equation*}
\sigma_B^2(k) = \frac{[\mu_T\omega(k) - \mu(k)]^2}{\omega(k)[1 - \omega(k)]}.
\end{equation*}

\subsection{Statistical quartiles}
\label{sec:quartile}

A quartile is a statistical term used when an ordered range is divided into four parts and the Boxplot chart, generally employed in data analyses, is a suitable visualization that uses such information~\cite{Frigge:1989}.
Thus, we have the first quartile $Q_1$ representing the number that limits the first 25\% values, the second quartile $Q_2$ limits the 50\% of the range (the median), and the third quartile $Q_3$ limits the 75\% of the range.
Furthermore, it is possible to calculate a dispersion measure named interquartile range $IQR = Q_3 - Q_1$ that can be used to identify the limit range values.
Almost 99.3\% of the range stays between the lower limit ($Q_1 - 1.5\times IQR$) and the higher limit ($Q_3 + 1.5\times IQR$).
The remaining values (outside the range) are named as outliers.

\section{Denoising method updated}
\label{sec:material_method}

We implemented two adaptations in the~\autoref{al:filter}.
First, the noise profile (lines 4-11) is defined by the automatic threshold selection as defined in~\autoref{sec:otsu}.
We assume there are two modes on a given frequency band, the foreground sounds and the background noise, and that a proper noise profile $M$ is the value that better describe the separation between these modes, what could generate a refined profile to distinct situations.

With the tool described in~\autoref{sec:quartile}, we remove outlier values from the band frequencies to avoid their impact into the process.
We iteractively identify and remove the values beyond lower and higher limits, until no outlier is encountered.
Then, we generate the histogram of the remaining values and also apply a moving average (\autoref{al:filter}, lines 6-7), and finally search for the threshold as described in~\autoref{sec:otsu}.
In this case, $bin$ represents real numbers, instead of integers between $[0, 255]$ as in image context.

The second modification is on the definition of the $\theta$ parameter, since it depends on previous knowledge of the audio set and it can be complex to tune.
Hence, we applied the previous outliers exclusion and threshold search to the average values $avg$ calculated by~\autoref{al:filter} (lines 18-19), and used it as $\theta$ parameter.
Finally, we also rescale the signal \textbf{s} to $[-1, 1]$ range before the filtering process and after the reconstruction of the signal we rescale the result to the original signal range.
That rescale improved filtering of low volume sounds.

\section{Background noise index}
\label{sec:bgn}

With the same process used to remove noise in~\autoref{sec:material_method}, we updated the Background noise index (BGN).
The noise profile generated is the spectral version of the BGN (named as BGN$_{sp}$) and the summary index BGN is the application of the process (removing outliers and searching threshold) to the Hilbert amplitude envelope of the audio signal.
Before the calculation, as described by~\citet{Towsey:2017}, we also convert values to decibel and truncate them to -90 dB when they are lesser than this value (truncation can not be necessary).

\section{Experimental results}
\label{sec:results}

This section presents some tests with our implementations.
We considered two clips with 30 seconds from two recordings with distinct attributes and collected under different approaches, one audio file from a terrestrial area and another from an underwater area.
Both audio files are from the dataset used by~\citet{Dias:2021}.

\begin{figure}[!h] 
  \centering    
  \begin{subfigure}[t]{0.5\textwidth}
  \centering    
  \includegraphics[scale=0.32]{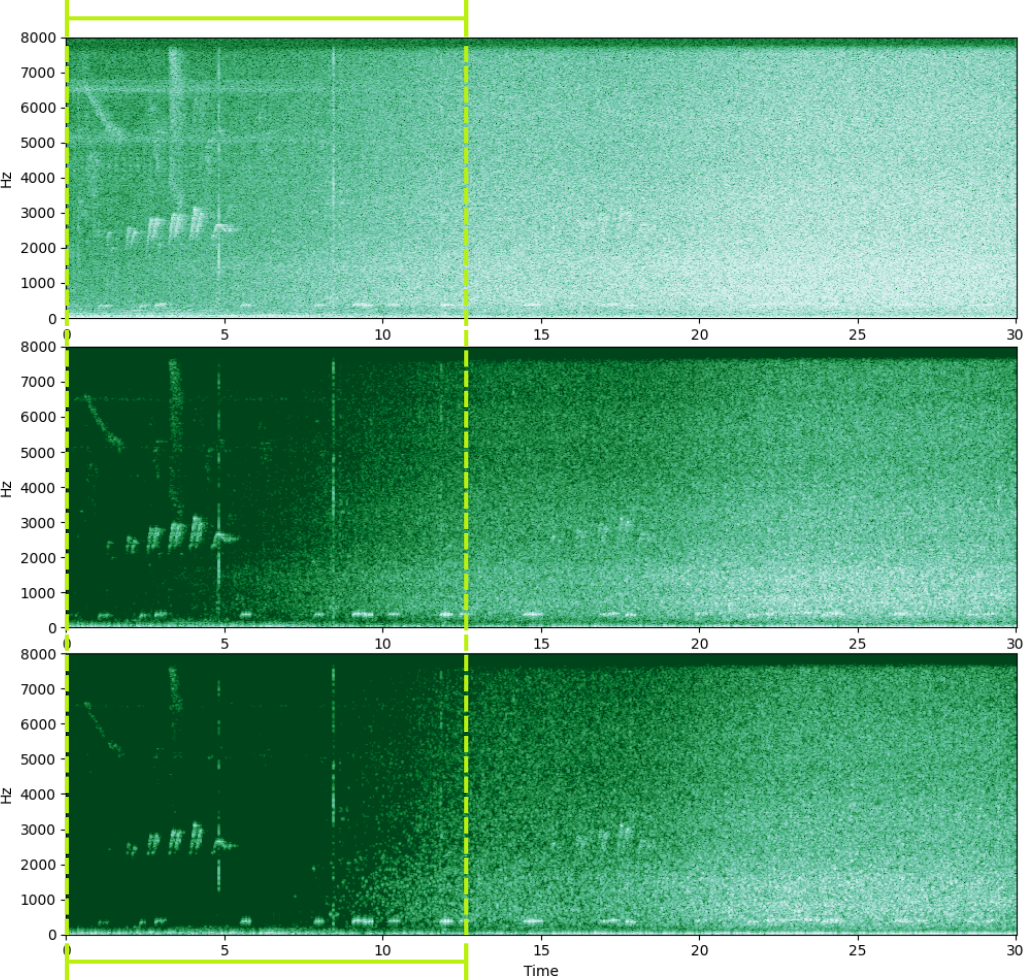} 
  \caption{ Spectrogram results } 
  \label{fig:costarica_spec} 
  \end{subfigure}~
  \begin{subfigure}[t]{0.5\textwidth}
  \centering    
  \includegraphics[scale=0.32]{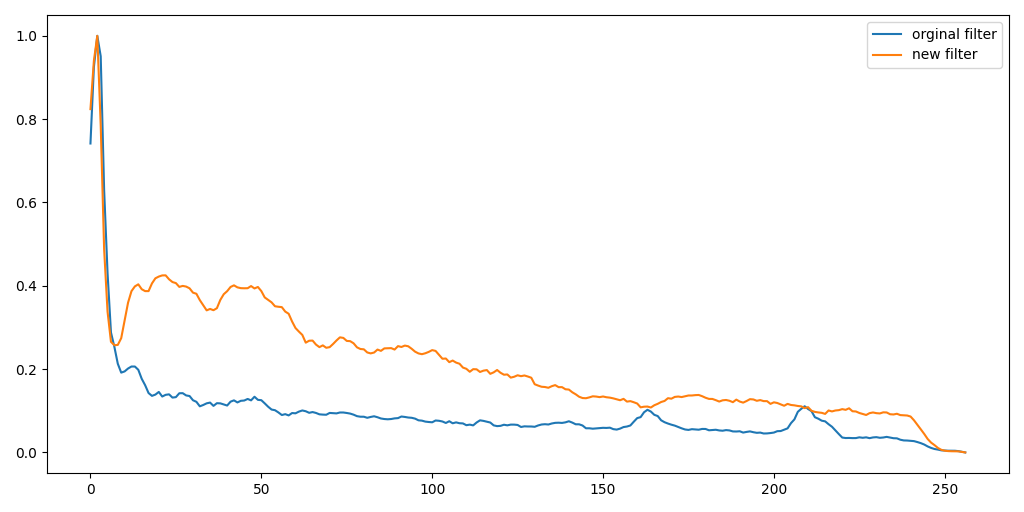} 
  \caption{ Noise profiles of filters } 
  \label{fig:costarica_profile} 
  \end{subfigure}  
  \caption{(a) Spectrograms of the terrestrial recording (top), 
  its filtered version (middle) with the original filter, and its filtered version (bottom) with the new filter.
  In the highlighted area (up to 12 seconds), the differences between the results of the filters are more perceptible.
  (b) Noise profile (BGN$_{sp}$) generated by the filter versions. Values were normalized to present curves in a suitable view}  
\end{figure}

\begin{figure}[!h] 
  \centering    
  \includegraphics[scale=0.5]{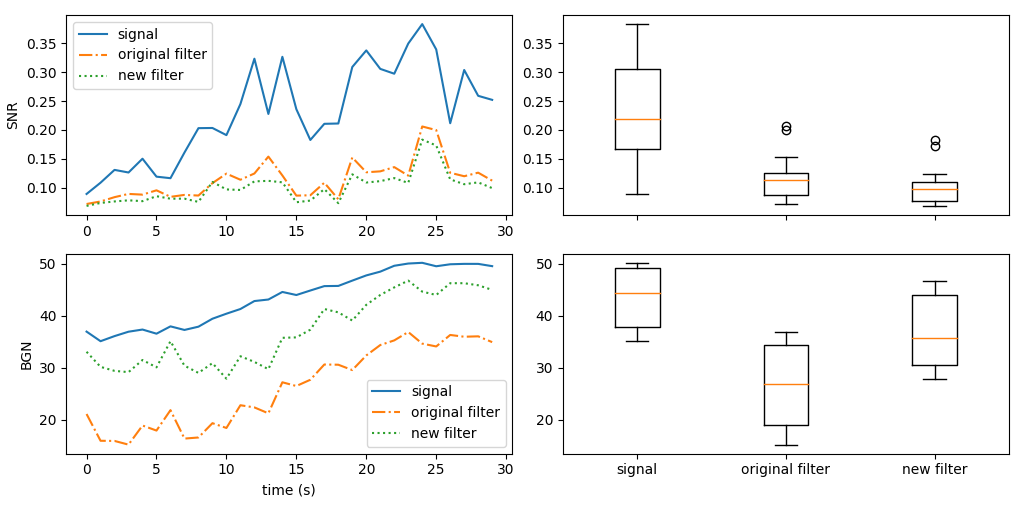} 
  \caption{Acoustic features of the terrestrial sound signal and the filtered signal with filter versions. Values calculated for each file second}
  \label{fig:costarica_indices} 
\end{figure}

\autoref{fig:costarica_spec} presents a terrestrial recording and its respective filtered versions (original and new filter). The recording has mainly biophony sounds, such as birds and insects, and geophony sounds such as heavy rain starting after 5 seconds.
There are great differences between filtered and non-filtered signals and the distinction between filter versions is easiest to
verify up to 12 seconds, where our approach removed more rain noise.
In~\autoref{fig:costarica_profile}, the profile yielded by the new filter has a larger range, which can explain the better cleaning presented with the spectrograms.
We also analyzed the sound to noise ratio (SNR)~\cite{Bedoya:2017} and the BGN index (see~\autoref{sec:bgn}), both calculated for each second of the recordings.
SNR is calculated as the relation between the mean and standard deviation of the power spectral density (PSD)~\cite{Welch:1967} of the audio signal.
Curves in~\autoref{fig:costarica_indices} follow the same ascending patterns and the new filter generated a signal with SNR mean lesser than the original filter (0.12 and 0.13, respectively).
Related to BGN, the new filter attained a greater mean than the original filter (37.27 and 26.44, respectively).

\begin{figure}[!h] 
  \centering    
  \begin{subfigure}[t]{0.5\textwidth}
  \centering    
  \includegraphics[scale=0.32]{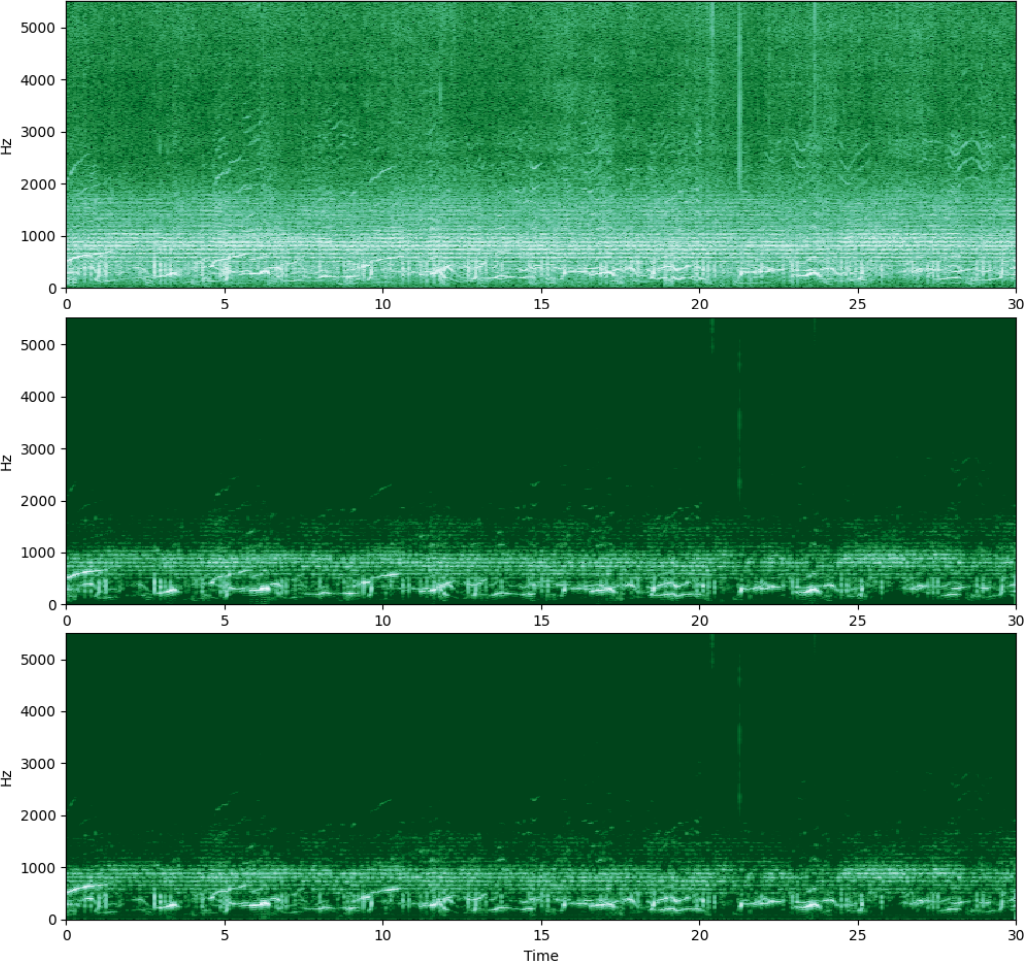} 
  \caption{ Spectrogram results } 
  \label{fig:ilheus_spec} 
  \end{subfigure}~
  \begin{subfigure}[t]{0.5\textwidth}
  \centering    
  \includegraphics[scale=0.32]{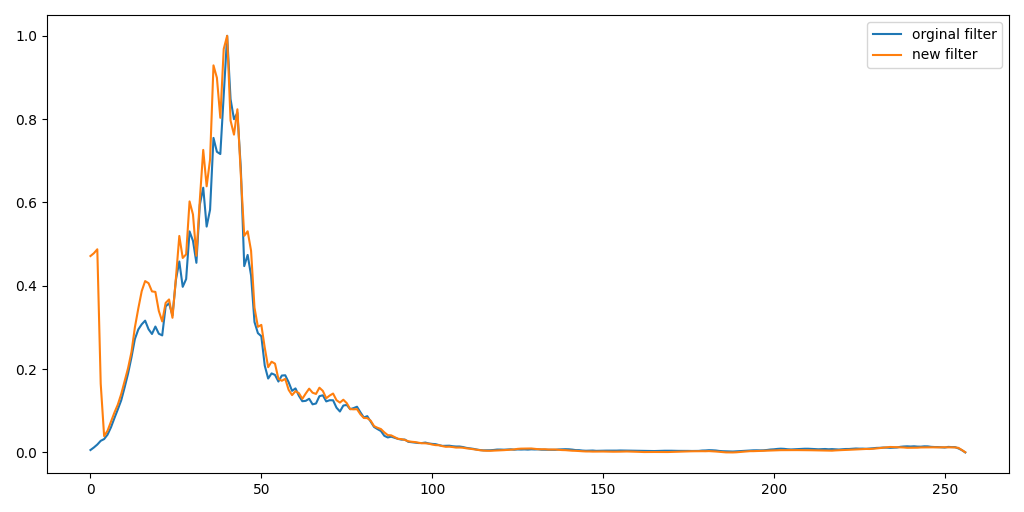} 
  \caption{ Noise profiles of filters } 
  \label{fig:ilheus_profile} 
  \end{subfigure}  
  \caption{(a) Spectrograms of the underwater recording (top), 
  its filtered version (middle) with the original filter, and its filtered version (bottom) with the new filter.
  (b) Noise profile (BGN$_{sp}$) generated by the filter versions. Values were normalized to present curves in a suitable view}    
  \label{fig:ilheus}  
\end{figure}

\begin{figure}[!h] 
  \centering    
  \includegraphics[scale=0.5]{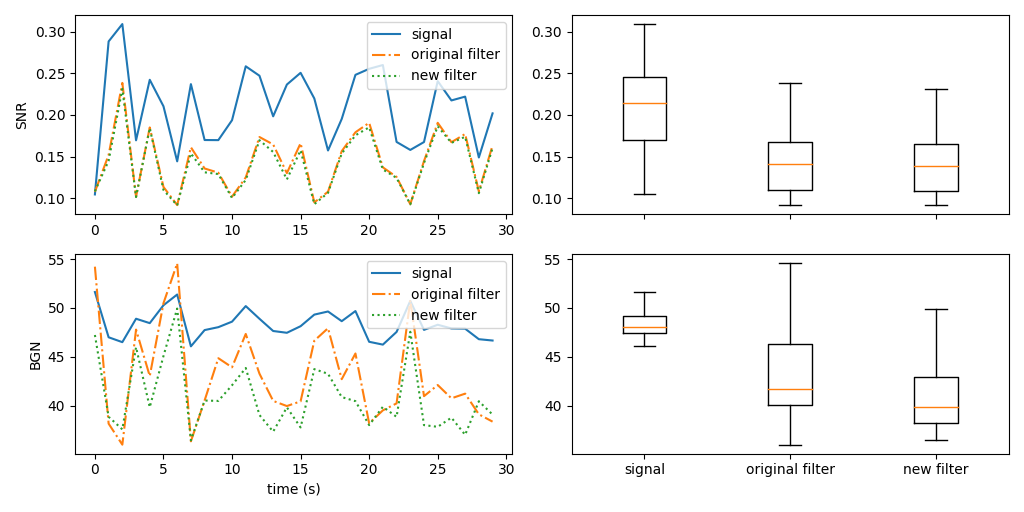} 
  \caption{Acoustic features of the underwater sound signal and the filtered signal with filter versions. Values calculated for each file second} 
  \label{fig:ilheus_indices} 
\end{figure}

On the other hand, filters results in the underwater recording are very similar, which can be verified with the inspection of~\autoref{fig:ilheus} and~\autoref{fig:ilheus_indices}.
That recording contains mainly humpback whale and fish choruses.
Curves in~\autoref{fig:ilheus_indices} are monotonic, SNR mean of the signal with the new filter is lesser than with the original filter (0.1494 and 0.1513, respectively), and BGN mean of the signal with the new filter is also lesser than the signal BGN mean with the original filter (42.57 and 43.70, respectively).

\section{Final Remarks}

Following those simple tests, the new implementation did not damage the application of the spectral filter and enhanced the estimation of BGN values.
The new filter removed more noise (rain) of terrestrial recording than its older version because the generated \emph{noise profile} could better represent frequency noise related to stationary sound such as rain.
When applied to underwater recording, the new filter attained similar noise attenuation to its old version.
Constant sounds as fish chorus were preserved, probably owing to frequency variations and overlaps with whale vocalizations.
It is noteworthy that an approach developed to filter terrestrial sounds can be applied to other cases such as underwater. 

Finally, our modifications presented straightforward and suitable ways to automatically define and tune the algorithm parameters, maintaining or improving the denoising results.
Furthermore, we recommend tests with other foreground sounds and background noise to assess the algorithm's robustness.  

\section*{Acknowledgements}

This study was financed in part by the Coordenação de Aperfeiçoamento de Pessoal de Nível Superior - Brasil (CAPES) - Finance Code 001, FAPESP (grant 2019/07316-0) and CNPq (National Council of Technological and Scientific Development) grant 304266/2020-5.

\clearpage
\bibliographystyle{elsarticle-num-names}
\bibliography{references,references_mestrado}

\end{document}